# The BINP HLS to measurement vertical changes on PAL-XFEL building and ground

**Hyo-Jin Choi, Kwang-Won Seo, Kye-Hwan Gil, Seung-Hwan Kim, Heung-Sik Kang**

*Department of Accelerator, PAL-XFEL, Pohang, Korea*

PAL-XFEL, a 4[th] generation light source, is currently being installed and will be completed by December of 2015 so that users can be supported beginning in 2016. PAL-XFEL equipment should continuously maintain the bunch-to-bunch beam parameter (60Hz, Energy 10GeV, Charge 200pC, Bunch Length 60fs, Emittance X/Y 0.481mm/0.256mm rad) in order to supply the energy, flux and timing of stable photons in tests by beam line users. To this end, PAL-XFEL equipment has to be kept precisely aligned (Linear Accelerator +/- 100μm, Undulator +/- 50μm). As a part of the process for installing PAL-XFEL, a GPS-used surface geodetic network is being constructed for precise equipment measurement and alignment, and the installation of a tunnel measurement network inside buildings is in preparation; additionally, the fiducialization of major equipment is underway. After PAL-XFEL equipment is optimized and aligned, if the ground and buildings go through vertical changes during operation, misalignment (and tilt) of equipment including various magnets and RF structures will cause errors in the electron beam trajectory, which will lead to changes to the beam parameter. For continuous and systemic measurement of vertical changes in buildings and to monitor ground sinking and uplifting, the Budker Institute of Nuclear Physics (BINP) Ultrasonic-type Hydrostatic Levelling System (HLS) is to be installed and operated in all sections of PAL-XFEL for linear accelerator, insertion device (Undulator) and beam line. This study will introduce the operation principle, design concept and advantages (self-calibration) of HLS, and will outline its installation plan and operation plan.

Email: choihyo@postech.ac.kr

Fax: +82-54-278-1799

# I. INTRODUCTION

During the construction of the Egypt Pyramids between 2600 and 2480 B.C., water was poured into an animal's gut in order to measure the hydrostatic level and strings were used to measure the wire position in the survey and alignment process. The Hydrostatic Levelling System (HLS) and Wire Position System (WPS) are still in use to format the horizontal axis in construction. The position of an object was measured by the human eye in the past, but these days it is measured by an electrical sensor and the data are analysed by computer - thanks to the advances in electronic equipment [1]. Recent advances in laser technology have made surveying an area of several tens of μm possible by using laser tracker [2]. But in cases where changes of the horizontal axis are measured continuously long-time, HLS and WPS, which are more precision (<1μm) then laser tracker, should be used.

## *HISTORY OF HLS USED ON THE ACCELERATOR*

The HLS was developed by the Alignment and Geodesy (ALGE) group at the European Synchrotron Radiation Facility (ESRF) for long term monitoring and control of rapid realignment of the Storage Ring machine. The concept of the non-contact capacitive sensor developed at the ESRF for the monitoring of level differences in the ESRF storage ring has been considerably improved upon by the company FOGALE-Nanotech. ESRF announced the results of twelve years of experience in HLS operation and measurement [3]. Various types of HLS Sensors, including capacitive and ultrasonic sensors, have been developed and used. European Council for Nuclear Research (CERN) announced the results of the comprehensive testing of HLS Sensor and WPS Sensor in many forms [4]. The status and useful information about HLS and WPS used by many research institutes around the world is available at CLIC Pre-Alignment Workshop and International Workshop on Accelerator Alignment (IWAA).

## HLS REFERENCE: WATER PIPE

The most important thing about HLS is the water pipe which provides the measurement reference. As shown in Figure 1, water within the water pipe should have good fluidity even with changes in the surrounding environment such as changes in temperature and pressure in order to maintain the constant level of water in the water pipe. It's the only way to calculate the floor deformation accurately using the measurement of all HLS Vessel floors. In terms of the fluid behaviour, after investigating studies about the way of calculating the water pipe diameter which is most appropriate for the length of full-filled and half-filled water pipes and the consequential stabilization time of water oscillation, the half-filled water pipe was found to be measured accurately [5][6]. Some studies even show a comparison between the thermal deformation of the material of the water pipe according to changes in temperature and changes in water volume [7]. To minimize the effects of thermal deformation, the length of the HLS sensor support should be shortened. The large-diameter PAL-XFEL should be used in order to provide the smooth flow and consistent level of water in the water pipe as shown in Figure 2.

Fig. 1. The surrounding environment influencing HLS.

Fig. 2. Half-filled water pipe.

## WATER VOLUME CHANGING BY TIDAL EFFECTS

The strength of gravity of planets in the solar system follows Isaac Newton's law of gravity and the superposition principle. There are three elements changing the gravity of the earth: the earth's orbit, the moon's orbit and leaning of the earth's rotational axis. As shown by Figure 3, the orbit of the earth circling around the sun is oval. The sun's tide generating force changes according to positions of the earth's orbit.

Fig. 3. Earth's orbit.

As shown by Figure 4, the effects of tide generating force appear due to complex movements, such as the earth's rotation, the moon's orbit and the leaning of the moon's orbital plane. It is very difficult to gain theoretical access to them and they can be various depending on factors (such as composition of the continental ground, latitude, longitude and altitude) affecting regions whose tide generating force is to be measured, so it is difficult to analyze them. The effects of the tide generating force changing over time lead to changes in gravity and in consequence the earth's land and sea affected by gravity display tidal phenomena over time.

Fig. 4. Distribution of the tidal force on the earth.

The tide generating force can be measured using an earth tide meter or a gravity meter. Figure 5 shows changes in the tide generating force measured in Korea using an earth tide meter. Changes in the tide generating force cause changes in the water volume and they appear as changes in the water height inside a water pipe in the process of measuring HLS. As shown by Figure 1, water produces volume changes because of various outside effects in addition to the tide generating force. In the case of water in a glass, the water height changes about 2μm/deg/cm because of temperature and the figure is about Max. 0.6μm/cm because of tides. There should be no temperature changes in order to observe the water height changing by tidal effects.

Fig. 5. The tide generating force depending on positions of the sun and the moon.

Although the water volume inside a water pipe dynamically changes moment by moment due to temperature inside a tunnel and tides, the water height of the entire area of the water pipe will be maintained in the short term if the flow inside the water pipe is large enough. The space inside the water pipe is closed and hydrostatic levelling measurement is made under the condition where the amount of water is the same even if the water volume changes. As shown by Figure 6, water pipes are installed on the floor inside the tunnel. As long as an accelerator works, entries to the tunnel are prevented and there is no vibration caused by people.

Fig. 6. The position of water pipes inside the accelerator tunnel.

There should be sufficient water flow for all of the water inside the water pipes to maintain the same water surface. Figure 7 shows the diameter of a water pipe to secure proper flow in accordance with the length of the pipe [5].

Fig. 7. The critical (optimum) depth of water and the period of oscillation for different lengths of pipes in a half-filled system.

Figure 8 shows the length and diameter of water pipes to be installed in PAL-XFEL. The material of pipes is anti-corrosive stainless steel SUS304 whose surface was treated sanitarily. The right diameter of a pipe for the length of the pipe was calculated according to Figure 7 and then pipes whose diameter is close to the calculation result were selected among pipes commercially available.

Fig. 8. The length and diameter of a water pipe.

## II. EXPERIMENTS AND DISCUSSION

*HLS DATA MEASUREMENT AND ANALYSIS*

To detect any errors generated in the installation process of HLS equipment ($\Delta H_{install}$) and resulting differences in floor height ($\Delta H_{floor}$), the absolute height of all HLS installed by laser tracker should be measured and all measurements of HLS sensor should be calibrated in a HLS data acquisition computer as shown in Figure 9. Such a measurement of absolute values should be performed once or twice a year because it takes long time.

Fig. 9. HLS absolute calibration.

After the calibration of all HLS errors through absolute value measurement, all HLS measurements should be expressed in relative values based on a reference HLS measurement value as shown in Figure 10. The level of water in the water pipe continually changes due to changes in the surrounding environment and the HLS reference measurement also changes due to ground motion. Changes in the ground motion of the reference HLS measurement should be measured once or twice a monthly by the laser tracker and calibrated by HLS data acquisition computer.

Fig. 10. HLS relative measurement (methods for routine measurement and displaying values).

HLS measurements include information such as ground level and various noise components affected by the environment (refer to Figure 1). The accurate ground change can be understood when the various noise components are reduced by analysis program. Figure 11 shows an example of long term measurements of the SPring-8 and the results of decomposition by the computer program, BAYTAP-G (Bayesian Tidal Analysis Program). This program is a method for tidal analysis proposed by Ishiguro based on the concept of Bayesian statistical modelling, Akaike's Bayesian Information Criterion (ABIC). BAYTAP-G decomposes the input data into (a) a tidal part, (b) a temperature part, (c) an irregular part, (d) a drift part under the assumption that the drift is smooth. The analysis model is obtained by maximizing the posterior distribution of the parameters. For the given data ABIC is used to select the optimum values of the hyper-parameters of the prior distribution and combination of parameters [6][8].

Fig. 11. BAYTAP-G Analysis of SPring-8.

*COMPOSITION OF BINP HLS AND PRINCIPLES OF MEASURING IT*

In the United States at 1914, R.A. Fessenden developed a moving coil transducer and succeeded in detecting an iceberg a mile away. After Paul Langevin succeeded in detecting a submarine 1500m away using the vacuum tube amplifier in France at 1918, Sound Navigation and Ranging (SONAR) began to be used widely. Aided by increased understanding of the characteristics of sound wave medium and development of sensor and electronics during two world wars, it is widely used for medical ultrasound testing and non-destructive testing.

A measurement concept of BINP HLS measuring the height of water using an ultrasonic transducer is shown in Figure 12. The transducer is H10KB3T (7MHz) used for ultrasonic flaw detectors made by GE Sensing & Inspection Technologies. The range that can be measured by the transducer is the far-field area. A reflector that adopts the role of an absolute ruler and the height of water surface can be measured correctly only when they are placed within the far-field area [9]. The height of the HLS bracket shown in Figure 8 should be adjusted properly so that the height of the water surface doesn't veer from the far field due to vertical changes in the building foundation. If there are serious vertical changes in the foundation, the water pipe support of Figure 8 should be adjusted too.

Fig. 12. The HLS measurement concept using an ultrasonic transducer.

As shown by Figure 13, ultrasonic waves that take place in the transducer are reflected in the reflector and water surface and are conveyed to the transducer. Even when the performance of the transducer and water temperature change, the height of the water surface can be measured correctly because the time gap between Wave-t1 and Wave-t2 reflected by the absolute ruler (D1) is 7.5mm. Such a self-calibration function improves the accuracy and credibility in the measurement of BINP HLS.

Fig. 13. Measuring the height of water surface using ultrasonic waves.

Figure 14 shows a block diagram, an electronic circuit of BINP HLS. Reflected ultrasonic waves are recorded as time at the TDC (Time to digital converter) through a comparator. Timing jitters can occur at the system clock 8MHz of an electronic circuit or time delays can occur at a microcontroller, but these effects are equally applied to all return waves. As shown by Figure 13, timing jitter and time delay elements of the electronic circuit are removed with a formula $[(t3-t1)/(t2-t1)]$ for calculating the length of D2 [10].

Fig. 14. Block diagram of the ULSE Electronics.

The resolution for measuring the distance of HLS is determined by the sound velocity in water and TDC of an electronic circuit. The sound velocity is determined by the temperature of water and the time resolution of TDC is determined by a system clock 8MHz. Like Figure 15, the resolution for measuring the distance of BINP HLS is about $0.2\mu m$.

Fig. 15. Resolution for measuring the distance of BINP HLS.

# III. CONCLUSION

## *PAL-XFEL BUILDING FOUNDATION*

The purpose of installing HLS is to continuously survey the vertical changes of a building and its foundation and record any changes. To analyse and understand the results of HLS measurement, people should know about the conditions of the building and its foundation. Conditions regarding the creation of the foundation of a PAL-XFEL building are shown in Figure 16. After deciding to construct a building at an altitude of 62 meters, earth at the altitude of 62 meters or higher was removed completely. In order not to construct the building on a weak foundation, the earth of the weathered zone, a weak foundation, was removed completely. After this, the space of the removed weak foundation was replaced with concrete to maintain an altitude of 62 meters. It did not pour the foundation piles for the foundation for enhanced bearing capacity due to construction of a PAL-XFEL building on the bedrock. Transformation of the building floor is connected with subsidence and upheaval of the foundation. Zones where the foundation is expected to change vertically can be found through continuously measuring vertical changes of the building floor using HLS. Measurement data of HLS is used for aligning accelerators.

Fig. 16. Conditions of creating the PAL-XFEL foundation.

## *BINP HLS TEST ON PAL-XFEL*

Figure 17 shows the method of using BINP HLS and the result of testing ULSE 2 sets which was borrowed from BINP to learn about the operation. The tidal effect of the sun and moon could not be confirmed because of the changes in surrounding temperature (2.2 degrees). The tidal effect can be seen in HLS when the influence of surrounding temperature is less than the tidal effect.

Fig. 17. BINP ULSE Test on PAL-XFEL.

## ACKNOWLEDGEMENT

Thanks to Dr. A.G. Chupyra and Dr. M.N. Kondaurov of BINP for their technical support and cooperation on testing BINP HLS in PAL-XFEL.

Figure Captions.

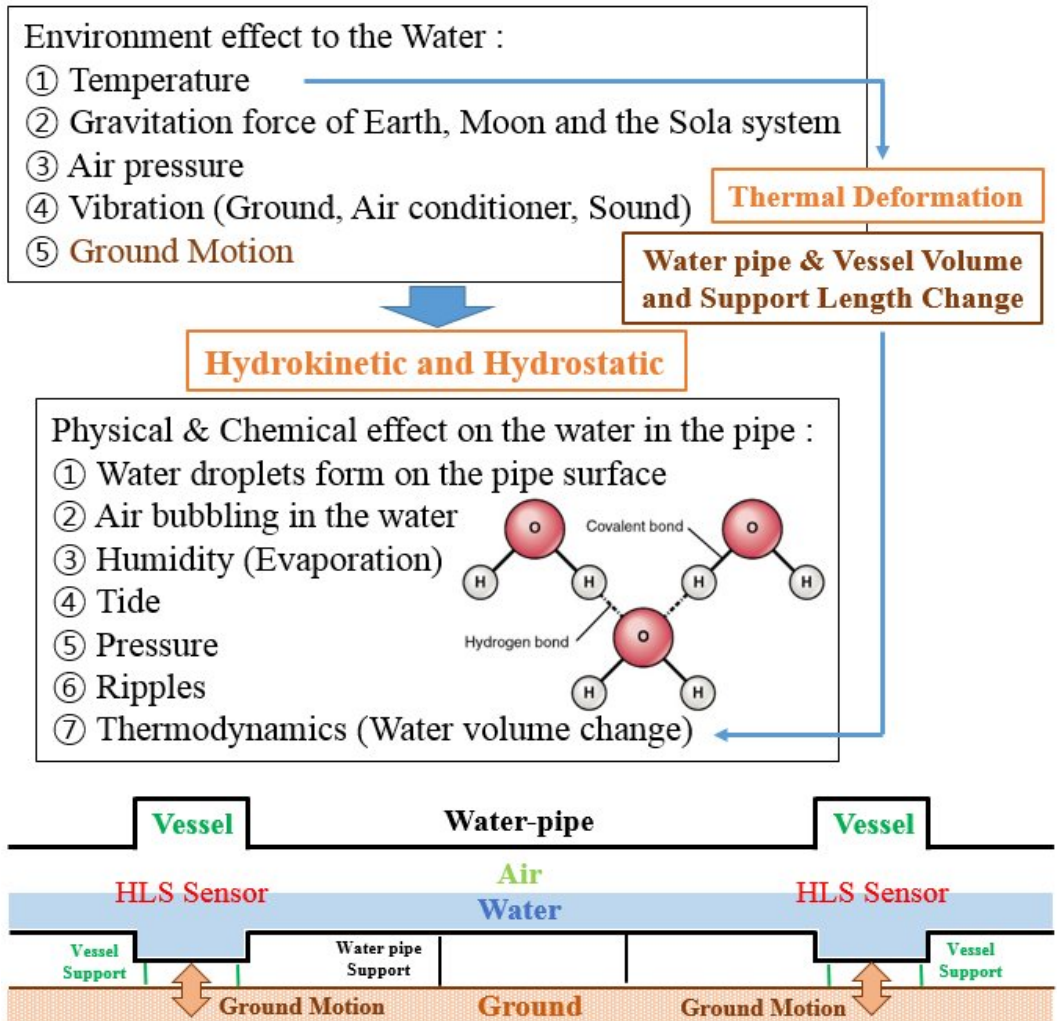

Fig. 1. The surrounding environment influencing HLS.

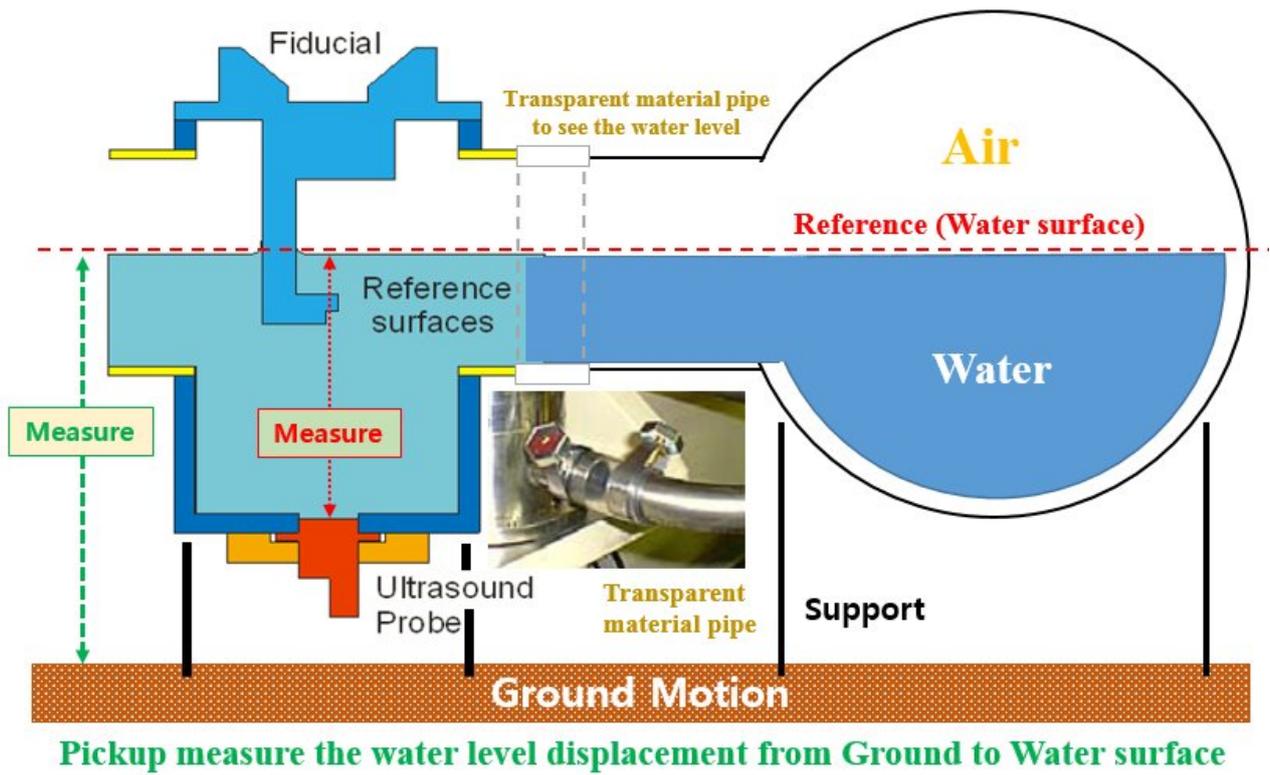

Fig. 2. Half-filled water pipe.

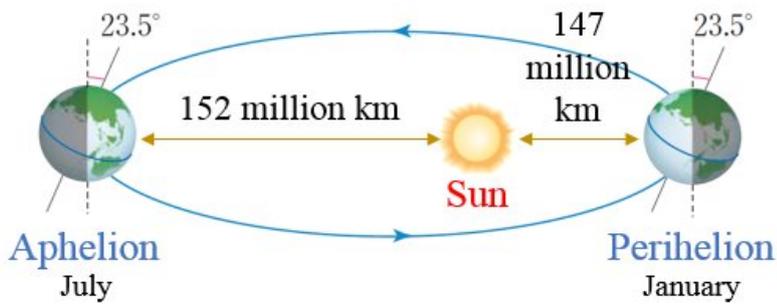

Fig. 3. Earth's orbit.

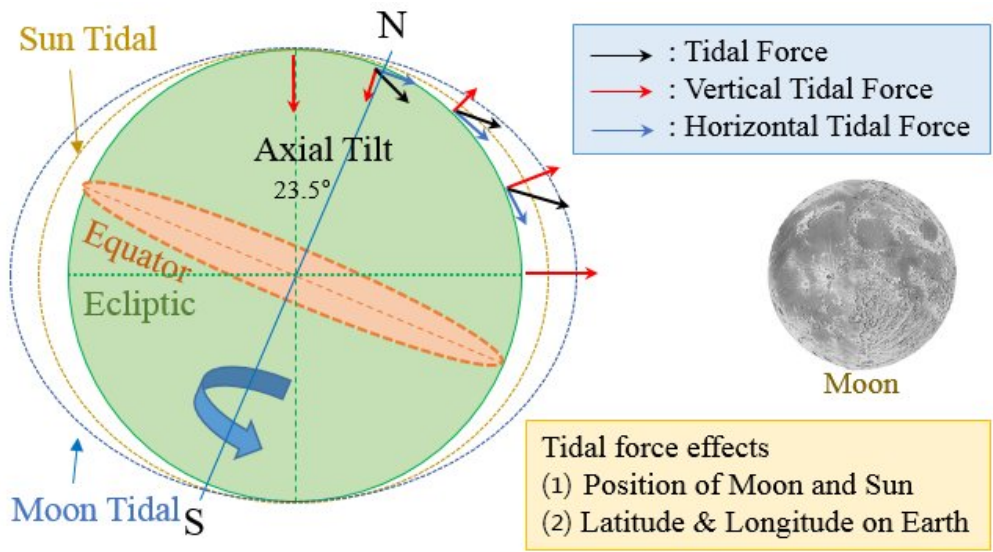

Fig. 4. Distribution of the tidal force on the earth.

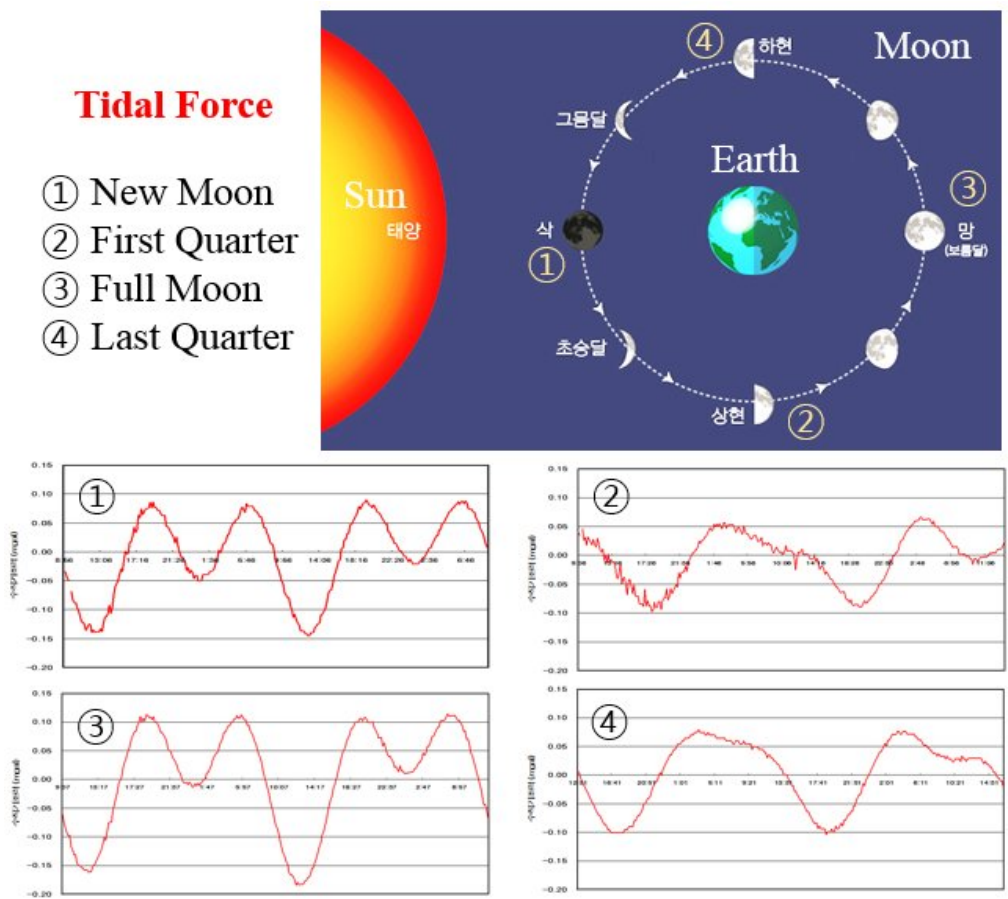

Fig. 5. The tide generating force depending on positions of the sun and the moon.

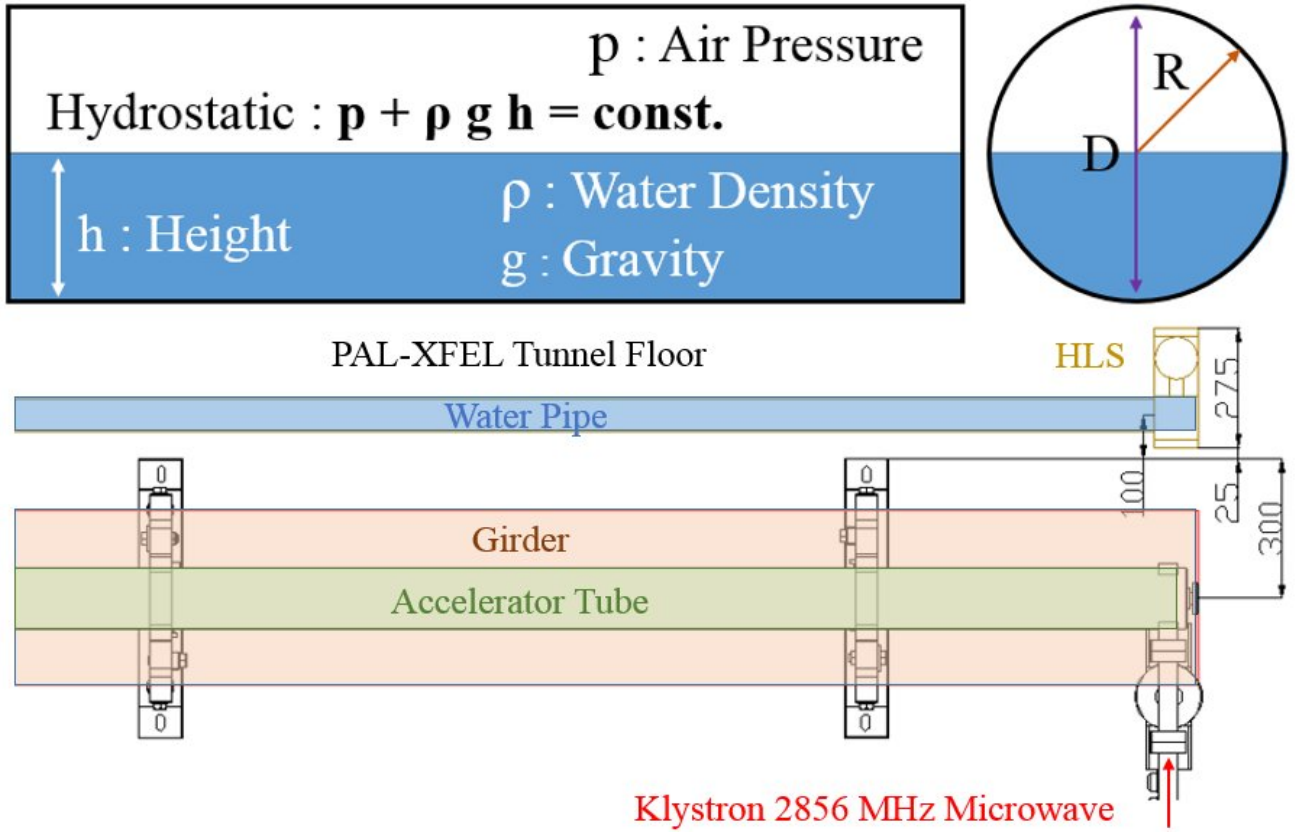

Fig. 6. The position of water pipes inside the accelerator tunnel.

| Distance (m) | Depth of water (m) | Period (sec) |
|---|---|---|
| 1 | 0.0028 | 12 |
| 5 | 0.0053 | 44 |
| 10 | 0.0070 | 76 |
| 50 | 0.0133 | 277 |
| 100 | 0.0175 | 484 |
| 500 | 0.0334 | 1752 |
| 1000 | 0.0440 | 3050 |

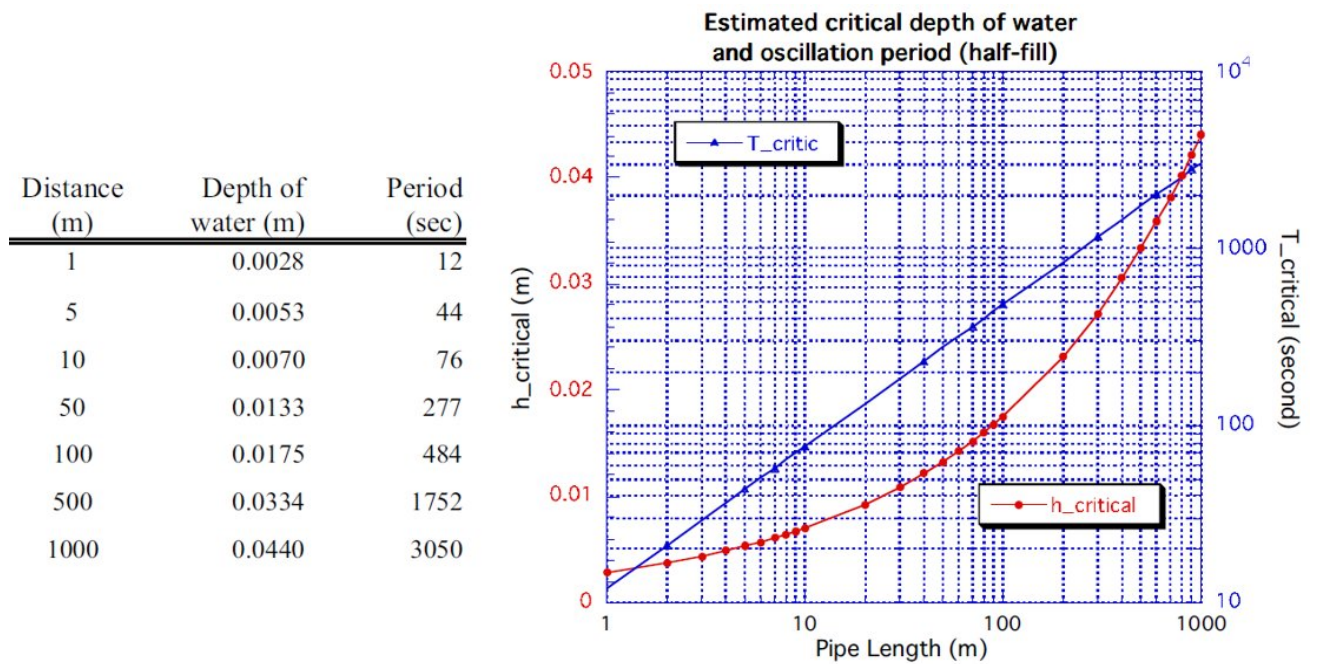

Fig. 7. The critical (optimum) depth of water and the period of oscillation for different lengths of pipes in a half-filled system.

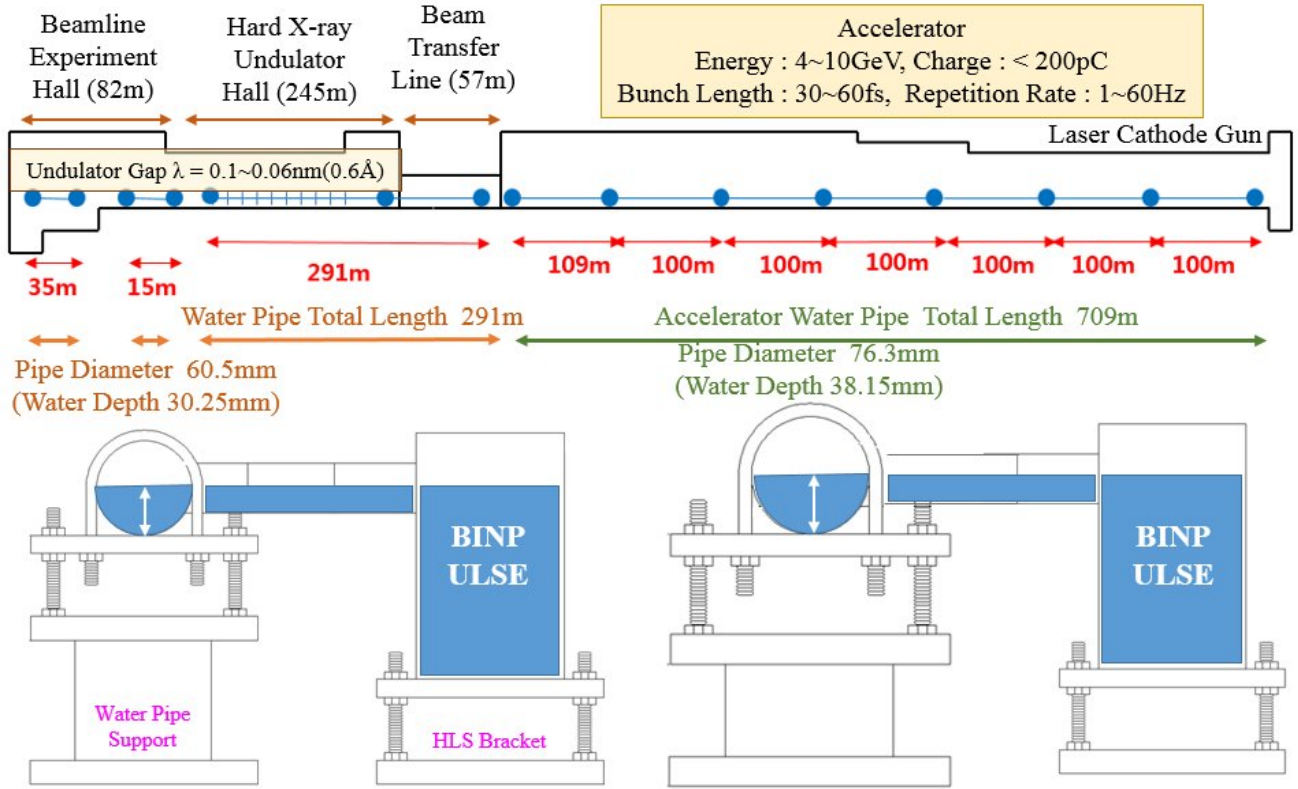

Fig. 8. The length and diameter of a water pipe.

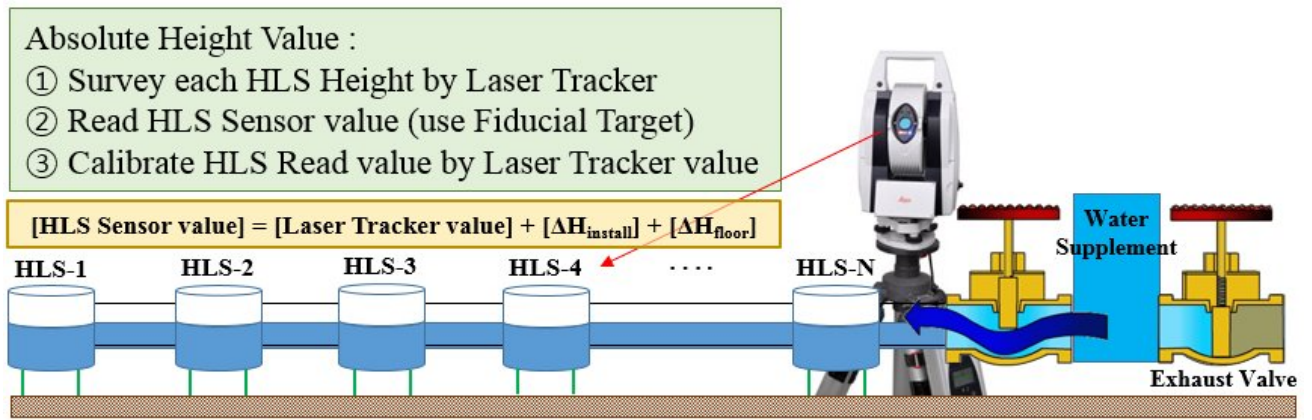

Fig. 9. HLS absolute calibration.

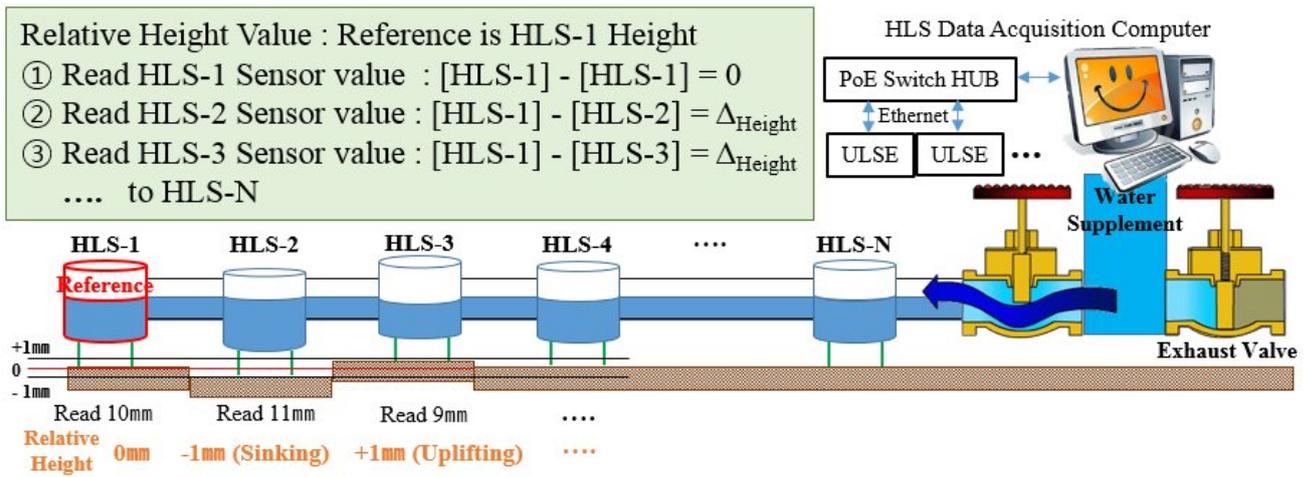

Fig. 10. HLS relative measurement (methods for routine measurement and displaying values).

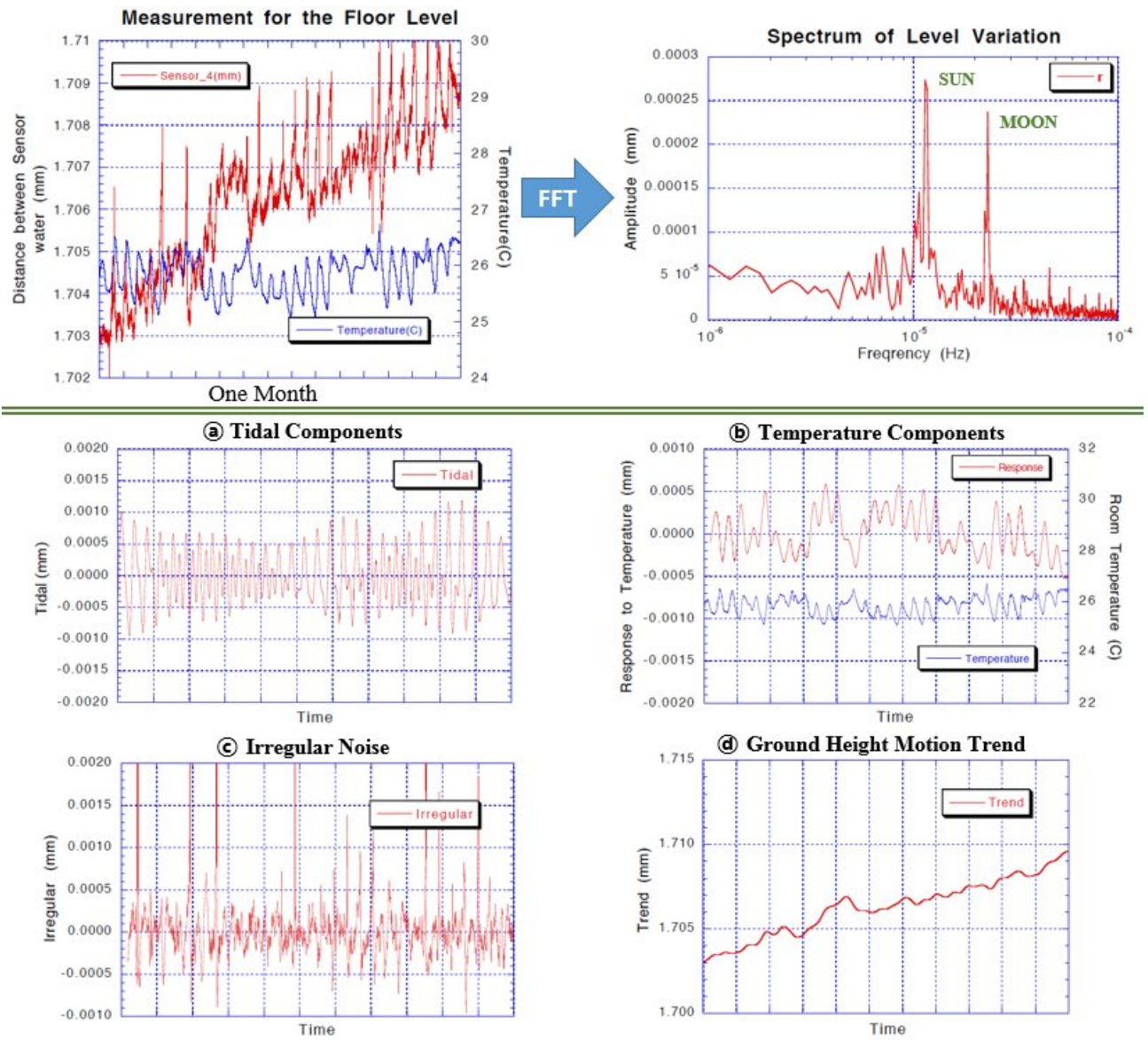

Fig. 11. BAYTAP-G Analysis of SPring-8.

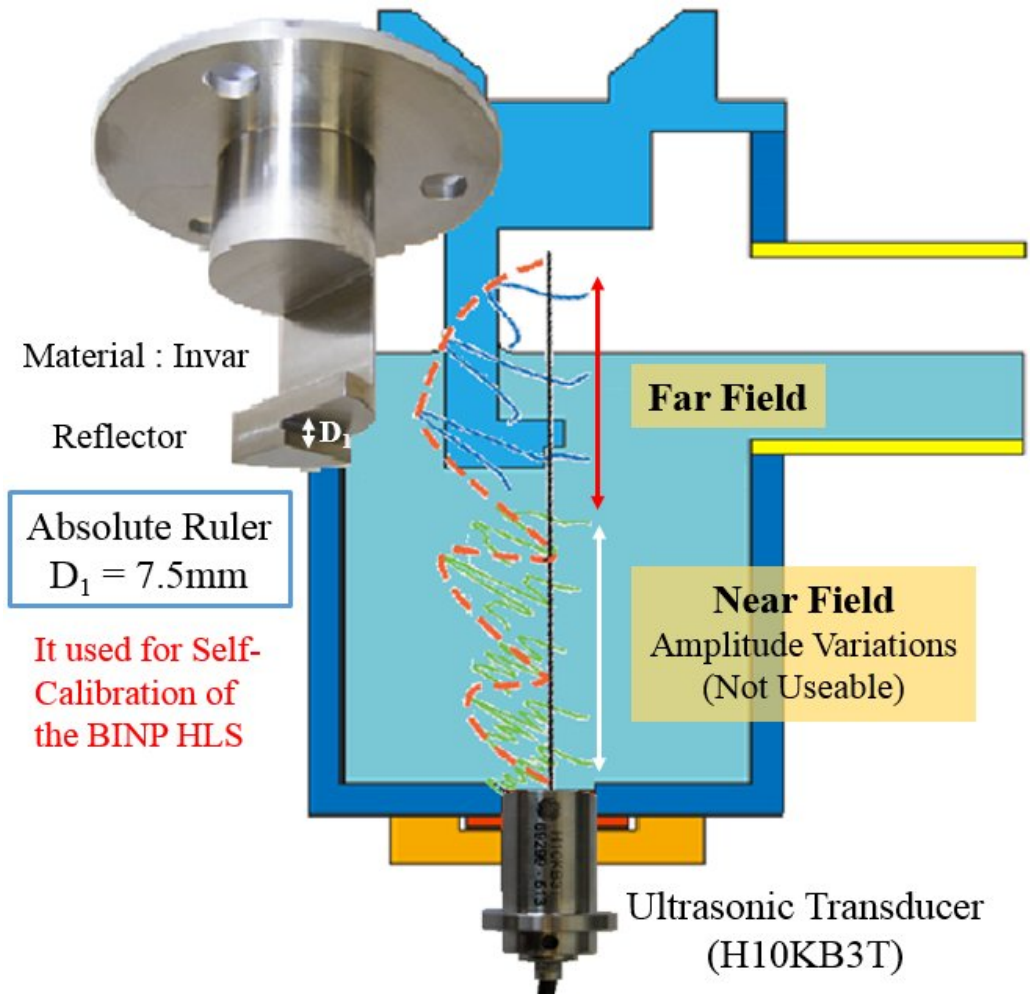

Fig. 12. The HLS measurement concept using an ultrasonic transducer.

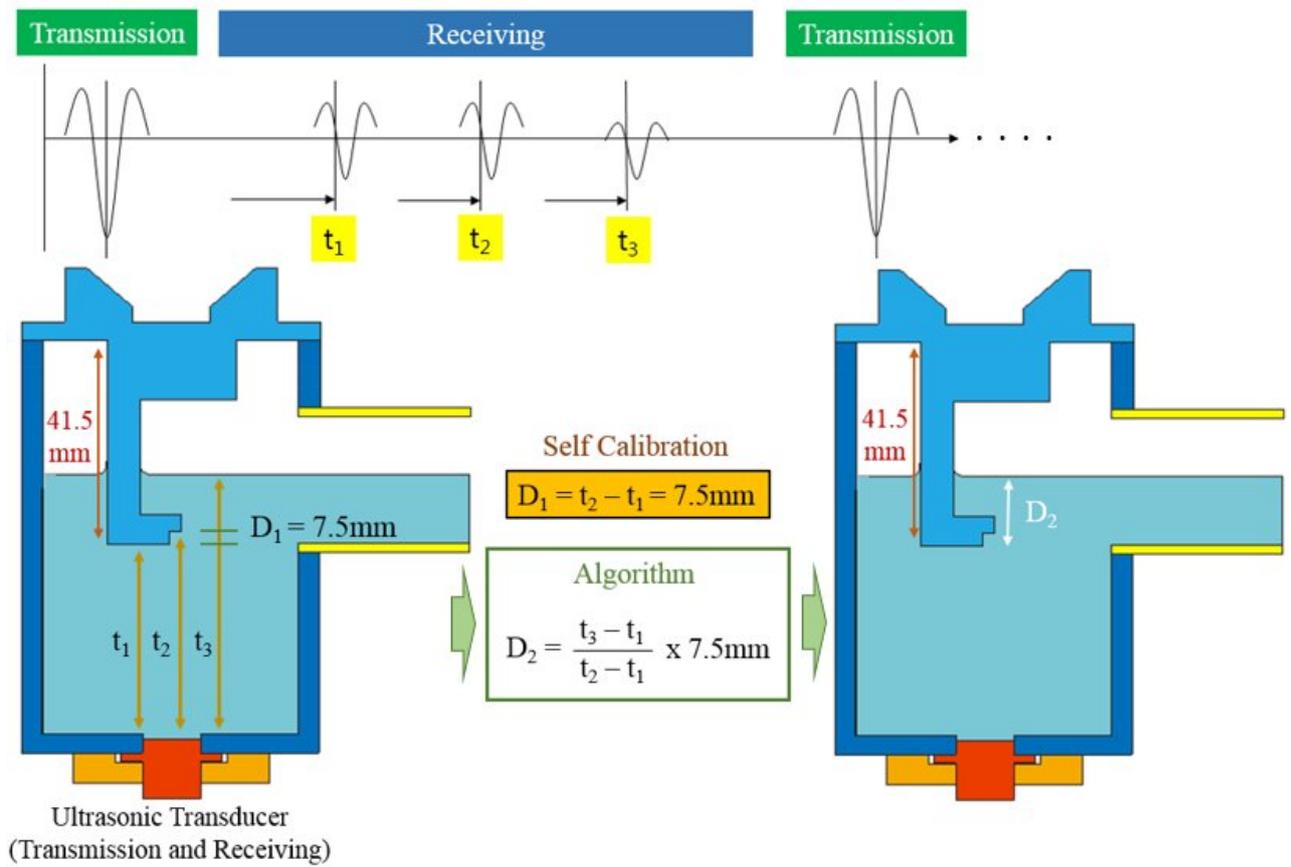

Fig. 13. Measuring the height of water surface using ultrasonic waves.

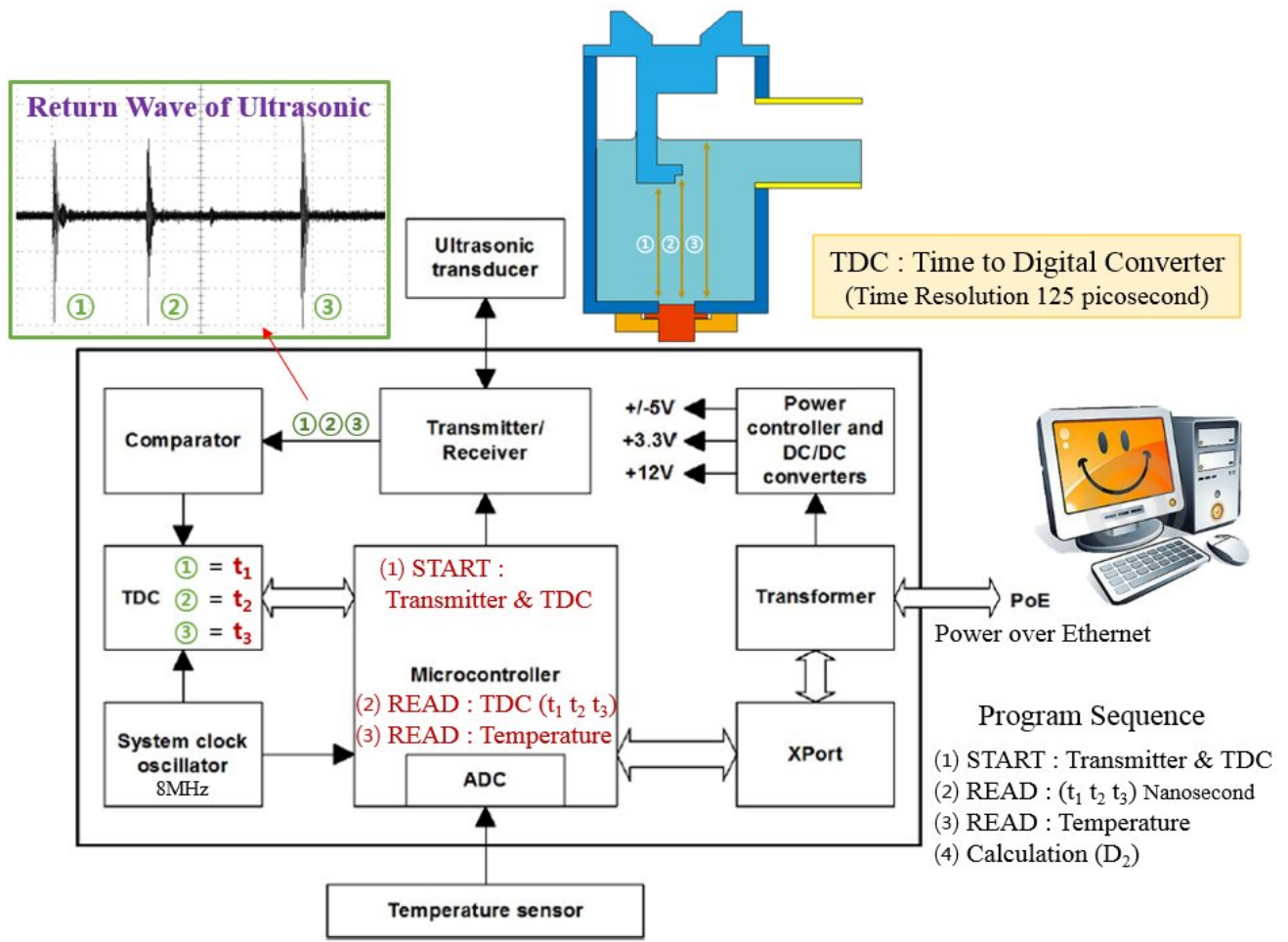

Fig. 14. Block diagram of the ULSE Electronics.

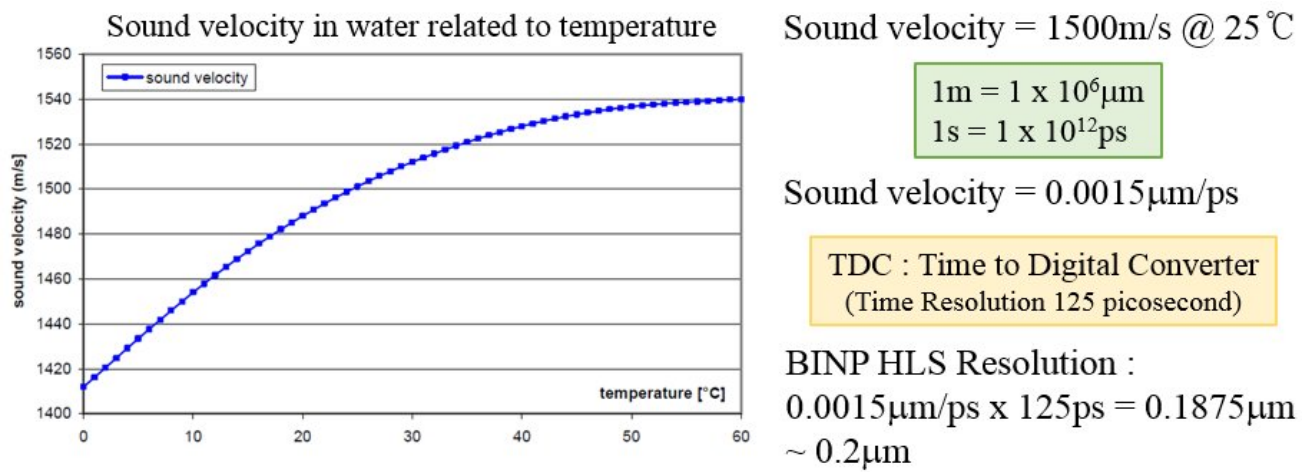

Fig. 15. Resolution for measuring the distance of BINP HLS.

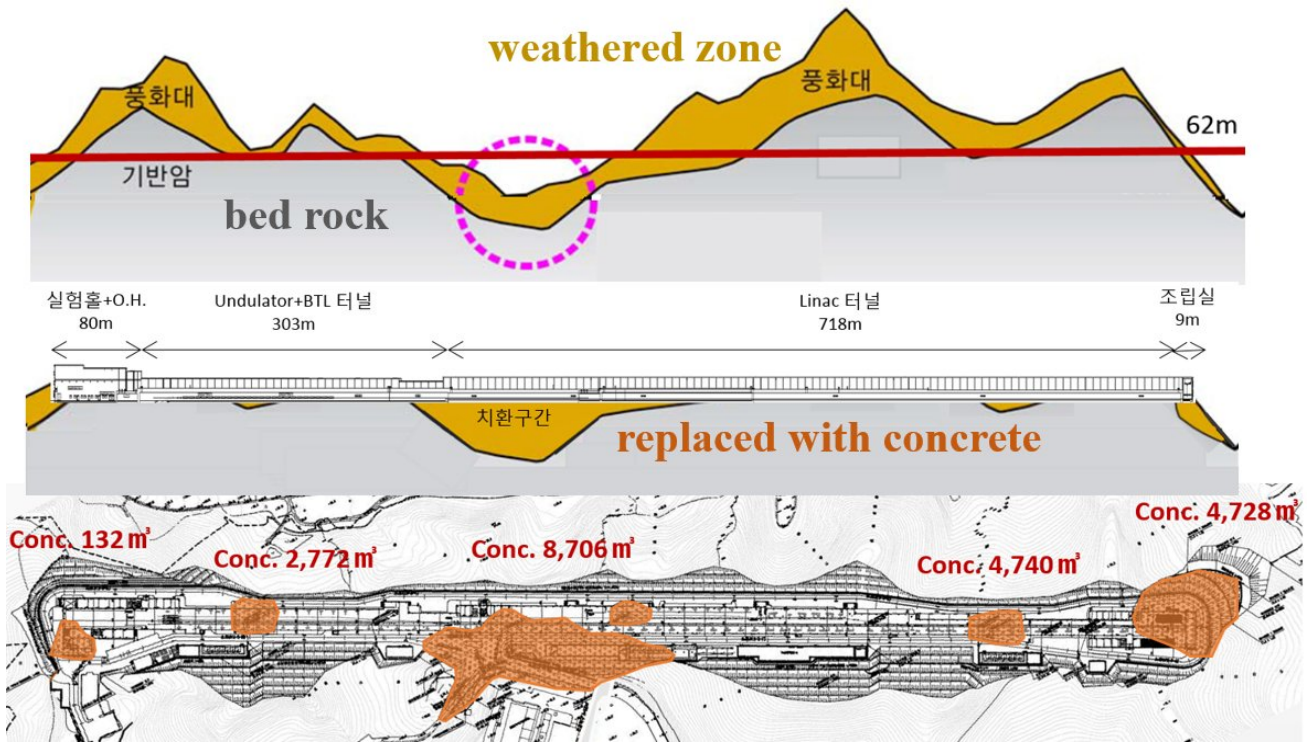

Fig. 16. Conditions of creating the PAL-XFEL foundation.

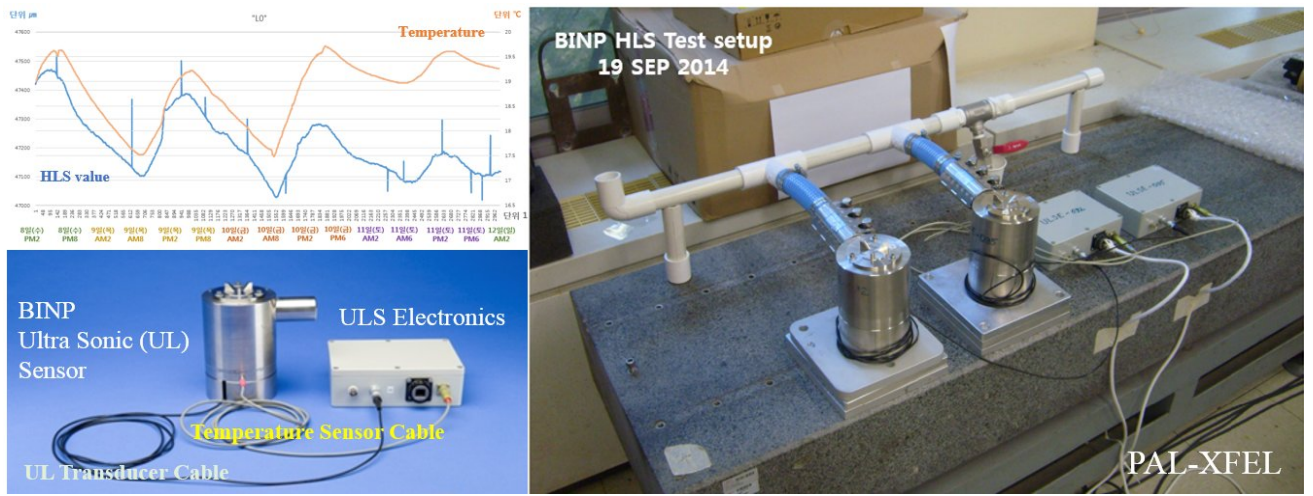

Fig. 17. BINP ULSE Test on PAL-XFEL.